\documentclass[aps,prb,reprint, amsmath]{revtex4-2}

\usepackage{graphics}
\usepackage{xcolor}
\usepackage{soul}

\definecolor{dartmouthgreen}{rgb}{0.05, 0.5, 0.06}

\begin{document}

\title{Piecewise Non-Linearity and Capacitance in the \\ Joint Density Functional Theory of Extended Interfaces}

\author{Tobias Binninger}
\email{tobias.binninger.science@gmx.de}
\affiliation{ICGM, Univ. Montpellier, CNRS, ENSCM, Montpellier, France}

\begin{abstract}
The \textit{ab initio} simulation of charged interfaces in the framework of density functional theory (DFT) is heavily employed for the study of electrochemical energy conversion processes. The capacitance is the primary descriptor for the response of the electrochemical interface. It is essentially equal to the inverse of the energy curvature as a function of electron number, and as such there appears a conflict with the fundamental principle of piecewise linearity in DFT that requires the energy curvature to be zero at fractional electron numbers, \textit{i.e.} almost everywhere. To resolve this conflict, we derive an exact expression between the energy curvature and the Kohn-Sham density of states, the local density of states, and the Fukui potential. We find that the piecewise linearity requirement does not hold for the volume- or area-specific energy of extended systems and surfaces. Applied to the joint density functional theory of an electrode--electrolyte interface, including the ionic and dielectric response of the electrolyte, the same expression represents a rigorous basis for the partitioning of the total interfacial capacitance into contributions of the quantum capacitance, space-charge capacitance, and electrochemical double-layer capacitance. It provides insight into the influence of the electrode material, thickness, and temperature on the charging characteristics, as demonstrated by results for a bulk gold electrode, a single-layer gold electrode, and a single-layer graphene electrode.
\end{abstract}

\maketitle

Since the work of Perdew \textit{et al.}~\cite{1982_PhysRevLett_Perdew}, piecewise linearity has become a cornerstone of the density functional theory (DFT) of open systems~\cite{1965_PhysRev_Mermin}. The ground-state energy $E$ as a function of the electron number $N$ consists of linear segments between integer values of $N$. In particular, the energy curvature $\eta = \partial^2E/\partial N^2$ is zero at fractional $N$. This property is explained by the statistical ensemble character of open system states~\cite{1982_PhysRevLett_Perdew}.  Within the Kohn-Sham (KS) approach~\cite{1965_PhysRev_Kohn_Sham}, the interacting many-electron system is mapped onto a system of non-interacting fermions comprising a series of single particle eigenstates $\epsilon_{i}$ with occupation numbers $f_{i}$ according to a Fermi-Dirac distribution. In the zero temperature limit and for a fractional electron number $N$, the Fermi energy is pinned to the energy $\epsilon_{\text{H}}$ of the highest (partially) occupied orbital (HOMO), and $\epsilon_{\text{H}} = \partial E/\partial N$ according to Janak's theorem~\cite{1978_PhysRevB_Janak}. Therefore, piecewise linearity is equivalent to a constant HOMO energy, which is fundamentally important for the construction of improved DFT functionals~\cite{2018_PhysRevX_Nguyen_Marzari} and the physical interpretation of KS orbital energies~\cite{1983_PhysRevLett_Perdew, 2017_PNAS_Perdew}. 

Conceptual DFT~\cite{1984_JACS_Parr_Yang} provides a different perspective on energy curvature, giving it a physical meaning as the chemical hardness $\eta = \partial^2E/\partial N^2$~\cite{1983_JACS_Parr_Pearson, 1985_PNAS_Yang_Parr} quantifying chemical reactivity. However, because of piecewise linearity, the definition of chemical hardness as a second derivative of the energy is problematic, and it must be replaced by its finite difference equivalent~\cite{1983_JACS_Parr_Pearson} involving integer electron numbers only. Obviously, such definition is only meaningful for systems with a finite number of electrons. In the limit of infinitely extended systems, the piecewise linearity condition is trivially fulfilled, because the system properties, and thus the KS orbital energies, become insensitive to finite changes in electron number~\cite{1983_PhysRevLett_Perdew}, at least if the added charge gets delocalized across the entire system~\cite{2008_PhysRevLett_Mori-Sanchez}. To define a non-trivial quantity, Vl\v{c}ek \textit{et al.}~\cite{2015_JChemPhys_Vlcek_Baer} introduced the energy curvature per unit cell $\eta_{\text{UC}} = \partial^2E_{\text{UC}}/\partial N_{\text{UC}}^2$ for periodic systems, where $E_{\text{UC}} = E_M/M$ and $N_{\text{UC}} = N_M/M$ are the energy and electron number per unit cell of a large, but finite crystal comprising $M$ unit cells. Obviously, the total energy curvature $\eta_M = \partial^2E_M/\partial N_M^2$ scales like $\eta_M = \eta_{\text{UC}}/M$. Unlike $\eta_M$, it was found that $\eta_{\text{UC}}$ did not turn to zero with increasing system size, which was attributed to the failure of common approximate exchange-correlation (XC) functionals to correctly reproduce the expected zero unit cell energy curvature~\cite{2015_JChemPhys_Vlcek_Baer}. 

For electrochemical interfaces, however, it follows from the Lippmann equation~\cite{2010_book_Schmickler} that the unit cell energy curvature is essentially equal to the inverse of the interfacial capacitance per unit cell, and it should thus be strictly greater than zero, in contrast to the expectation from piecewise linearity. Moreover, the total interfacial capacitance is commonly split into contributions of the quantum capacitance~\cite{1985_JPhysChem_Gerischer, 2007_ApplPhysLett_Fang}, space-charge capacitance, and electrochemical double-layer capacitance, but their precise relation to fundamental DFT is not fully understood to date~\cite{2011_EnergyEnvSci_Stoller, 2015_PhysRevB_Radin_Wood, 2016_JPhysChemLett_Zhan, 2020_JElectroanalChem_Schmickler}. For most systems, explicit modelling of the electrolyte~\cite{2018_JChemPhys_Gross, 2019_JPhysChemLett_Sprik} within DFT is prohibitively expensive, and different implicit models for the ionic counter charge in the electrolyte  were developed, including a homogeneous background~\cite{2006_AngewandteChem_Filhol}, Gaussian distributions~\cite{2019_JChemPhys_Hoermann_Marzari}, and diffuse screening layers described by Poisson-Boltzmann-type equations~\cite{2006_PhysRevB_Otani, 2008_PhysRevB_Jinnouchi, 2012_PhysRevB_Letchworth-Weaver_Arias, 2017_JChemPhys_Sundararaman_Arias, 2019_JChemPhys_Mathew_Hennig, 2019_JChemPhys_Nattino_Marzari,
2019_JChemPhys_Melander_Honkala}. The latter approach emerges from a general joint density functional theory (JDFT)~\cite{2005_JPhysChemB_Petrosyan_Arias, 2012_PhysRevB_Letchworth-Weaver_Arias} that combines the electronic DFT with a description of the electrolyte in terms of dielectric solvent and ionic charge densities. Obviously, the electrolyte model will influence capacitance and energy curvature~\cite{2018_JChemPhys_Schwarz}.

In the present letter, we consolidate the various perspectives on energy curvature. We show that the piecewise linearity requirement breaks down for unit cells of extended periodic systems, and we derive an exact expression that links the total energy curvature to the density of states and Fukui functions of the system. It is valid for finite and for extended systems, both metals and insulators. Including the influence of an electrolyte environment within JDFT, it further provides a natural partitioning of the total capacitance into quantum capacitance, XC-capacitance, space-charge capacitance, and electrolyte capacitance. Applied to bulk metal electrodes, a particularly intuitive expression involving the Fukui surface dipole is obtained. 

Mermin~\cite{1965_PhysRev_Mermin} formulated the grand canonical version of density functional theory for electrons in a fixed external potential $(-e)\phi_{\text{ext}}(\mathbf{r})$. In a typical situation, the latter is the electrostatic potential $\phi_{\text{ext}}(\mathbf{r}) = (1/4\pi\epsilon_0)\int\rho_{\text{ext}}(\mathbf{r}')/|\mathbf{r}-\mathbf{r}'|\text{d}\mathbf{r}'$ produced by an external charge density $\rho_{\text{ext}}$ that represents the atomic cores. For given temperature $T$ and chemical potential $\mu$, the grand potential $\Omega = U-TS-\mu N$, where $U$ and $S$ are the inner energy and entropy, respectively, is a functional of the electron density $n(\mathbf{r})$ with a minimum for the equilibrium density. The self-consistent Kohn-Sham solution~\cite{1965_PhysRev_Kohn_Sham} to the minimization problem is given by 
\begin{align}
\label{eq_KS_electron_density}
n(\mathbf{r}) = \sum_i \omega_i \,n_i(\mathbf{r})\ ,
\end{align}
where $\omega_i = 1/\left(1+\exp\left[(\epsilon_i-\mu)/kT\right]\right)$ are the Fermi-Dirac occupation numbers, and $n_i(\mathbf{r}) = |\Psi_i(\mathbf{r})|^2$ are the normalized densities of the eigenstates of the non-interacting single particle Schr\"odinger equation,
\begin{align}
\label{eq_KS_Schroedinger}
\left(-\frac{\hbar^2}{2m}\nabla^2 + (-e)\phi(\mathbf{r}) + \mu_{\text{xc}}(\mathbf{r}) \right)\Psi_i(\mathbf{r}) = \epsilon_i\,\Psi_i(\mathbf{r}) \ .
\end{align}
The XC-potential $\mu_{\text{xc}}(\mathbf{r}) = \delta F_{\text{xc}}/\delta n(\mathbf{r})$ is the variational derivative of the XC-energy functional, and $\phi(\mathbf{r}) = \phi_{\text{ext}}(\mathbf{r}) - (e/4\pi\epsilon_0)\int n(\mathbf{r}')/|\mathbf{r}-\mathbf{r}'|\text{d}\mathbf{r}'$ is the sum of the external and electronic electrostatic potentials, where $\epsilon_0$ is the vacuum permittivity. Integrating Eq.~\eqref{eq_KS_electron_density} over space and using the normalization of the orbital densities $n_i$ yields the (average) electron number
\begin{align}
\label{eq_electron_number_Fermi}
N = \sum_i \frac{1}{1+\exp\left(\frac{\epsilon_i-\mu}{kT}\right)} \ .
\end{align}

We first note that, by construction, the chemical potential $\mu$ in the Fermi-Dirac distribution of the KS orbital occupations $\omega_i$ is equal to the chemical potential of the electronic grand canonical ensemble, 
\begin{align}
\label{eq_chemical_potential_free_energy}
\mu = \frac{\partial A}{\partial N} \ ,
\end{align}
with the Helmholtz free energy $A = \Omega + \mu N$. This relation can be regarded as Janak's theorem in grand canonical DFT. Consequently, chemical hardness is given by the free energy curvature
\begin{align}
\label{eq_hardness_free_energy}
\eta = \frac{\partial^2A}{\partial N^2} = \frac{\partial \mu}{\partial N} \ .
\end{align}
Taking the partial derivative of Eq.~\eqref{eq_electron_number_Fermi} w.r.t. $N$ and resolving for $\partial\mu/\partial N$, we obtain
\begin{align}
\label{eq_hardness_I}
\frac{\partial \mu}{\partial N} = \frac{1}{g^T_{\text{D}}(\mu)}\left(1 + \sum_i p^T(\mu-\epsilon_i)\frac{\partial \epsilon_i}{\partial N}\right) \ ,
\end{align}
where we introduced the temperature-dependent density of states (DOS) $g^T_{\text{D}}(\epsilon) = \sum_i p^T(\epsilon-\epsilon_i)$, with $p^T(x) = (1/kT)\,\exp[x/kT]/\left(1+\exp[x/kT]\right)^2$ being a thermally broadened peak with unit area centered at zero. Obviously, $p^T(x) \rightarrow \delta(x)$ for $T\rightarrow 0$, and therefore $g^T_{\text{D}}(\epsilon) \rightarrow g_\text{DOS}(\epsilon)$, which is the usual DOS of the KS spectrum. Using the Hellmann-Feynman theorem with the Hamiltonian of Eq.~\eqref{eq_KS_Schroedinger}, we compute
\begin{align}
\label{eq_KS_eigenval_deriv}
& \frac{\partial \epsilon_i}{\partial N} = \int |\Psi_i(\mathbf{r})|^2 \left((-e)\frac{\partial \phi(\mathbf{r})}{\partial N} + \frac{\partial \mu_{\text{xc}}(\mathbf{r})}{\partial N}\right) \text{d}\mathbf{r} \\
& = \iint n_i(\mathbf{r}) \left(\frac{e^2}{4\pi\epsilon_0}\frac{1}{|\mathbf{r}-\mathbf{r}'|} + \frac{\delta \mu_{\text{xc}}(\mathbf{r})}{\delta n(\mathbf{r}')}\right) f_{\text{e}}(\mathbf{r}')\text{d}\mathbf{r}'\text{d}\mathbf{r} \ , \nonumber
\end{align}
where $\delta \mu_{\text{xc}}(\mathbf{r})/\delta n(\mathbf{r}')$ is the XC-kernel, and $f_{\text{e}}(\mathbf{r}')=\partial n(\mathbf{r}')/\partial N$ is the electronic Fukui function~\cite{1984_JACS_Parr_Yang}, which fulfills $\int f_{\text{e}}(\mathbf{r}') \text{d}\mathbf{r}' = 1$. Inserting Eq.~\eqref{eq_KS_eigenval_deriv} into~\eqref{eq_hardness_I} yields the free energy curvature, or chemical hardness,
\begin{align}
\eta\, =\, \frac{1}{g^T_{\text{D}}(\mu)}\, + \int \frac{g^T_{\text{LD}}(\mu,\mathbf{r})}{g^T_{\text{D}}(\mu)} \left\{\mu^f_{\text{xc}}(\mathbf{r})-e\phi^f_{\text{e}}(\mathbf{r})\right\}\text{d}\mathbf{r} \ ,
\label{eq_hardness_II}
\end{align}
where $g^T_{\text{LD}}(\epsilon,\mathbf{r}) = \sum_i p^T(\epsilon-\epsilon_i)n_i(\mathbf{r})$ is the temperature- dependent local density of states (LDOS), which fulfills $\int g^T_{\text{LD}}(\epsilon,\mathbf{r}) \text{d}\mathbf{r} = g^T_{\text{D}}(\epsilon)$ and converges to the usual LDOS $g_\text{LDOS}(\epsilon,\mathbf{r})$ for $T\rightarrow 0$. Further, $\mu^f_{\text{xc}}(\mathbf{r})=\int[\delta \mu_{\text{xc}}(\mathbf{r})/\delta n(\mathbf{r}')] f_{\text{e}}(\mathbf{r}')\,\text{d}\mathbf{r}'$ is the Fukui XC-potential, and $\phi^f_{\text{e}}(\mathbf{r})=(-e/4\pi\epsilon_0)\int f_{\text{e}}(\mathbf{r}')/|\mathbf{r}-\mathbf{r}'|\,\text{d}\mathbf{r}'$ is the electrostatic Fukui potential~\cite{2011_JPhysChemA_Cardenas}. 

Eq.~\eqref{eq_hardness_II} is a first important result of the present letter, and it is valid for any temperature. Yang and Parr~\cite{1985_PNAS_Yang_Parr} already obtained the first term, which neglects the KS orbital relaxation as pointed out by Cohen \textit{et al.}~\cite{1994_JChemPhys_Cohen}. This is generally termed the frozen orbital approximation in conceptual DFT, and the fixed band approximation (FBA) in quantum capacitance~\cite{2015_PhysRevB_Radin_Wood}. The second term in Eq.~\eqref{eq_hardness_II} is the interaction between the normalized LDOS $(g^T_{\text{LD}}/g^T_{\text{D}})$ at the Fermi energy and the Fukui function mediated by the kernel of the KS-potential. 

For a finite, confined system, the KS eigenvalue spectrum is discrete. At $T=0$, the DOS becomes a series of Dirac delta-functions, and the normalized LDOS at the Fermi energy converges precisely to the HOMO density $n_{\text{H}}(\mathbf{r})$. The chemical potential has discontinuous jumps at integer $N$, and for fractional $N$ it is pinned to the respective HOMO energy $\epsilon_{\text{H}}$, where the DOS is infinite. Therefore, the first term in Eq.~\eqref{eq_hardness_II} is zero at fractional $N$ and $T=0$ for a finite, confined system, and Eq.~\eqref{eq_hardness_II} becomes precisely equal to the energy curvature relation (8) in Vl\v{c}ek \textit{et al.}~\cite{2015_JChemPhys_Vlcek_Baer}, which must be zero to fulfill the piecewise linearity requirement. Accordingly, to provide a meaningful definition of the chemical hardness $\eta$ for finite systems, Eq.~\eqref{eq_hardness_free_energy} is generally replaced by its finite difference equivalent involving integer electron numbers only~\cite{1983_JACS_Parr_Pearson}.

We next consider an infinite periodic system as the limit $M \rightarrow \infty$ of finite cyrstals comprising $M$ unit cells (UC), and, following Vl\v{c}ek \textit{et al.}~\cite{2015_JChemPhys_Vlcek_Baer}, we consider the unit cell energy curvature $\eta_{\text{UC}} = M\eta_M$. From the $M$-scaling of the DOS, LDOS, and Fukui function, together with the unit-cell periodicity, it follows that $\eta_{\text{UC}}$ becomes constant for large $M$, and it fulfills an expression analogous to Eq.~\eqref{eq_hardness_II} involving the corresponding UC-related quantities. Most importantly, for an infinite system, $\eta_{\text{UC}}$ generally does not converge to zero with $T\rightarrow 0$, neither for gapless metals, nor for insulators with a KS band gap. With increasing $M\rightarrow\infty$, the discrete KS orbitals keep splitting and form continuous bands in the thermally broadened UC-normalized DOS $g^T_{\text{D}}(\epsilon)$ at any $T>0$. The subsequent limit $T\rightarrow 0$ yields a UC-normalized DOS which remains bounded everywhere, in particular $1/g_{\text{DOS}}(\mu)>0$. Also the second term of Eq.~\eqref{eq_hardness_II} generally remains non-zero if $\lim_{M\rightarrow\infty}$ is performed before $\lim_{T\rightarrow 0}$, because piecewise linearity only dictates the contribution of the actual HOMO to be zero, which is negligible compared to the contributions of all orbitals with infinitesimally smaller eigenvalues of the same band at any $T>0$. Consequently, the energy per unit cell of an infinite periodic system does not fulfill piecewise linearity, and the value of the energy curvature is determined by the properties of the band in which the chemical potential lies. For metals, $\eta=\partial\mu/\partial N$ is well-defined. For semiconductors or insulators with a gap above the HOMO, the derivative yields two different values when taken from above ($\eta^+$) and below ($\eta^-$), corresponding to the properties of the conduction band and valence band, respectively. 

Until now, we have neglected the question of charge neutrality. However, for infinite periodic systems, the unit cell must remain neutral upon change of the electron number to avoid a divergent electrostatic energy~\cite{2008_PhysRevB_Sharma}. For bulk systems, this is typically achieved by a homogeneous compensating background charge~\cite{2015_JChemPhys_Vlcek_Baer}. For electrochemical interfaces, charge neutrality is maintained by changing the ionic charge in the electrochemical double layer in accordance with the electron number. The total energy cannot be separated into uniquely defined electronic and ionic energies, because electrostatic energy contributions can be arbitrarily shifted from electronic to ionic, and \textit{vice versa}, by shifting the electrostatic potential reference. Consequently, the energy curvature, as well as the capacitance, must be understood as properties of the system as a whole.

JDFT~\cite{2005_JPhysChemB_Petrosyan_Arias, 2012_PhysRevB_Letchworth-Weaver_Arias} treats the ionic densities $\{n_i\}$ and the dielectric bound charge density $\rho_{\text{diel}}$ of the electrolyte together with the electron density $n$ in a combined free energy functional. We consider a large, but finite electrode--electrolyte system comprising $M$ supercells of an electrode slab embedded in a bulk of electrolyte that is sufficiently extended so that fields and charges are fully screened at the boundary of the system. The ionic charge changes with the electron number and preserves overall neutrality $Q_{\text{tot}}=\int (\rho_{\text{ext}}+\rho_{\text{e}}+\rho_{\text{ion}}+\rho_{\text{diel}})\,\text{d}\mathbf{r} = 0$, where $\rho_{\text{e}} = -e\,n$ and $\rho_{\text{ion}} = \sum eZ_i\,n_i$ for ion species with charges $eZ_i$. The charge neutrality constraint is automatically fulfilled for a simple electrolyte response described by a linearized Poisson-Boltzmann equation~\cite{2012_PhysRevB_Letchworth-Weaver_Arias, 2017_JChemPhys_Sundararaman_Arias}. For general electrolyte models, however, charge neutrality must be imposed~\cite{2013_ModelSimulMaterSciEng_Gunceler}. At a general level of JDFT~\cite{2012_PhysRevB_Letchworth-Weaver_Arias, 2017_JChemPhys_Sundararaman_Arias}, the total Helmholtz free energy functional can be written
\begin{align}
\label{eq_JDFT_free_energy}
& A\left[n,\{n_i\}, \rho_{\text{diel}}, \phi\right] \ =\ T^{\text{\,ni}}_{\text{kin}}[n] \,-\, TS^{\text{ni}}[n] \,+\, F_{\text{xc}}[n] \nonumber\\
& \qquad + A^0_{\text{elyte}}\left[\{n_i\}, \rho_{\text{diel}}\right] \,-\, \frac{\epsilon_0}{2}\int|\nabla\phi|^2\,\text{d}\mathbf{r} \nonumber\\
& \qquad + \int \phi\,\left(\rho_{\text{ext}}-e\,n+\sum eZ_i\,n_i+\rho_{\text{diel}}\right) \text{d}\mathbf{r} \ ,
\end{align}
where $T^{\text{\,ni}}_{\text{kin}}[n]$ and $S^{\text{ni}}[n]$ are the functionals of kinetic energy and entropy of the non-interacting Kohn-Sham fermions~\cite{2017_JChemPhys_Sundararaman_Arias}, respectively, $F_{\text{xc}}[n]$ is the electronic exchange-correlation functional, and $A^0_{\text{elyte}}\left[\{n_i\}, \rho_{\text{diel}}\right]$ is the free energy functional of the electrolyte excluding mean-field electrostatic interactions. All mean-field electrostatic interactions in the system are captured by the terms with the electrostatic potential $\phi$. Minimization w.r.t. variations $\delta\phi(\mathbf{r})$, i.e. setting $\delta A/\delta \phi(\mathbf{r}) = 0$, yields the Poisson equation $\nabla^2\phi=-(\rho_{\text{ext}}+\rho_{\text{e}}+\rho_{\text{ion}}+\rho_{\text{diel}})/\epsilon_0$ with solution
\begin{align}
\label{eq_electrostatic_potential_solution}
\phi(\mathbf{r}) = \frac{1}{4\pi\epsilon_0}\,\int \frac{\left[\rho_{\text{ext}}+\rho_{\text{e}}+\rho_{\text{ion}}+\rho_{\text{diel}}\right](\mathbf{r}')}{|\mathbf{r}-\mathbf{r}'|}\,\text{d}\mathbf{r}' \ .
\end{align}

We consider a constant temperature $T$ and volume $V$, and a controlled electron number $N$. We further treat the electrolyte as a reservoir with fixed ionic chemical potentials $\{\mu_i\}$. Under these conditions, the functional to be minimized is given by 
\begin{align}
\label{eq_free_energy_minimization}
& \mathcal{F}[n,\{n_i\}, \rho_{\text{diel}}, \phi] = A[n,\{n_i\}, \rho_{\text{diel}}, \phi] - \sum \mu_i\int n_i\,\text{d}\mathbf{r} \nonumber\\
& - \mu\left(\int n\,\text{d}\mathbf{r} - N\right) - \lambda \int\left(\rho_{\text{ext}}+\rho_{\text{e}}+\rho_{\text{ion}}+\rho_{\text{diel}}\right) \text{d}\mathbf{r} \ ,
\end{align}
where the Legendre transformations from ion numbers $N_i = \int n_i\,\text{d}\mathbf{r}$ to chemical potentials $\mu_i$ were performed. The electron chemical potential $\mu$ appears as the Lagrange multiplier of the electron number constraint, and $\lambda$ is the Lagrange multiplier of the charge neutrality constraint. Minimization w.r.t. the electron density yields $\mu = \delta A/\delta n(\mathbf{r}) + e\,\lambda$, or
\begin{align}
\label{eq_electronic_minimization}
\mu\ = \ \frac{\delta T^{\text{\,ni}}_{\text{kin}}}{\delta n(\mathbf{r})} - T\frac{\delta S^{\text{ni}}}{\delta n(\mathbf{r})} + \frac{\delta F_{\text{xc}}}{\delta n(\mathbf{r})} - e\,(\phi(\mathbf{r})-\lambda) \ ,
\end{align}
which is precisely solved~\cite{1965_PhysRev_Kohn_Sham} by the Kohn-Sham equations~\eqref{eq_KS_electron_density} and~\eqref{eq_KS_Schroedinger} with an electrostatic potential shifted by $-\lambda$. Minimization w.r.t. $\rho_{\text{diel}}$ results in $\delta A/\delta \rho_{\text{diel}}(\mathbf{r}) = \lambda$, and for the ionic densities we find $\mu_i = \delta A/\delta n_i(\mathbf{r}) - eZ_i\,\lambda$, or
\begin{align}
\label{eq_ionic_minimization}
\mu_i\ = \ \frac{\delta A^0_{\text{elyte}}}{\delta n_i(\mathbf{r})} + eZ_i\,(\phi(\mathbf{r})-\lambda)\ =\ \mu^0_{i,\text{b}} + eZ_i\,(\phi_{\text{b}}-\lambda) \ ,
\end{align}
where $\mu^0_{i,\text{b}} = \delta A^0_{\text{elyte}}/\delta n_{i,\text{b}}$ and $\phi_{\text{b}}=\phi(\mathbf{r}_\text{b})$ are the non-electrostatic parts of the ionic chemical potentials and the plateau value of the electrostatic potential~\eqref{eq_electrostatic_potential_solution} in the bulk electrolyte, respectively. The value of $\mu^0_{i,\text{b}}$ is determined by the fixed bulk concentrations of the electrolyte. Again, the Lagrange multiplier $\lambda$ is equivalent to a constant shift in the electrostatic potential and thus represents a gauge freedom. We note from Eq.~\eqref{eq_ionic_minimization} that the fixed ionic chemical potentials must be of the form $\mu_i = \mu^0_{i,\text{b}} + eZ_i\,\phi_0$, where $\phi_0$ is an arbitrarily chosen constant. It then follows that $\lambda = \phi_{\text{b}}-\phi_0$ with the consequence that the shifted potential $\phi(\mathbf{r})-\lambda$ is fixed at $\phi_0$ in the bulk electrolyte.

We now focus on equilibrium states of the system that fulfill the minimization conditions above. The minimized free energy~\eqref{eq_free_energy_minimization} of the system is then simply the grand potential w.r.t. the ionic variables,
\begin{align}
\label{eq_grand_potential_ionic}
\mathcal{F}(N,\{\mu_i\})\ =\ A - \sum \mu_i\,N_i \ ,
\end{align}
where we omitted the implicit dependence on temperature $T$ and volume $V$.
For fixed ionic chemical potentials $\{\mu_i\}$, the ion numbers $N_i$ change as a function of the electron number $N$ to preserve charge neutrality. Therefore, any change of $N$ implies a joint particle number change in the neutral electron--ion system. We thus define the derivative of the free energy~\eqref{eq_grand_potential_ionic} w.r.t. $N$ as the \textit{joint chemical potential}
\begin{align}
\label{eq_joint_chemical_potential}
\mu_{\text{J}} = \left(\frac{\partial \mathcal{F}}{\partial N}\right)_{\{\mu_i\}} = & \int\left(\frac{\delta A}{\delta n}\,\frac{\partial n}{\partial N} + \sum\left(\frac{\delta A}{\delta n_i} - \mu_i\right)\frac{\partial n_i}{\partial N} \right.\nonumber\\ 
& \quad\left. +\, \frac{\delta A}{\delta \rho_{\text{diel}}}\,\frac{\partial \rho_{\text{diel}}}{\partial N}\right)\text{d}\mathbf{r}\ =\ \mu \ ,
\end{align}
where we used the minimization conditions together with charge neutrality, $\partial Q_{\text{tot}}/\partial N = 0$. We note that $\mu_{\text{J}}$ is simply equal to the electron chemical potential $\mu$. The latter is determined by Eq.~\eqref{eq_electronic_minimization} together with $N = \int n\,\text{d}\mathbf{r}$, which are precisely equivalent~\cite{1965_PhysRev_Kohn_Sham} to the standard Kohn-Sham equations~\eqref{eq_KS_electron_density} and~\eqref{eq_KS_Schroedinger}, but with $\phi(\mathbf{r})$ in Eq.~\eqref{eq_KS_Schroedinger} replaced by the shifted electrostatic potential $\phi(\mathbf{r})-\lambda = \phi(\mathbf{r})-\phi_{\text{b}}+\phi_0$.

We can therefore follow the same derivation that led to Eq.~\eqref{eq_hardness_II} to obtain the total free energy curvature, or \textit{joint chemical hardness}
\begin{align}
& \eta_{\text{J}}\ =\ \frac{1}{g^T_{\text{D}}(\mu)} + \int \Gamma^T_{\text{LD}}\,\left\{\mu^f_{\text{xc}}-e\left[\phi^f_{\text{J}}-\phi^f_{\text{J},\text{b}}\right]\right\}\text{d}\mathbf{r}\ ,
\label{eq_hardness_JDFT}
\end{align}
where $\Gamma^T_{\text{LD}}(\mathbf{r})=g^T_{\text{LD}}(\mu,\mathbf{r})/g^T_{\text{D}}(\mu)$ is the normalized LDOS at the Fermi energy and we used the fact that $\phi_0$ is a constant, i.e. $\partial\phi_0/\partial N = 0$. According to Eq.~\eqref{eq_electrostatic_potential_solution}, the joint electrostatic Fukui potential $\phi^f_{\text{J}}(\mathbf{r})=\partial\phi(\mathbf{r})/\partial N=(-e/4\pi\epsilon_0)\int f_{\text{J}}(\mathbf{r}')/|\mathbf{r}-\mathbf{r}'|\,\text{d}\mathbf{r}'$ is generated by the joint Fukui function $f_{\text{J}} = f_{\text{e}}^{\text{J}}-f_{\text{ion+diel}}^{\text{J}}$, where we introduced an ionic and dielectric Fukui function $f_{\text{ion+diel}}^{\text{J}}=(1/e)\partial(\rho_{\text{ion}}+\rho_{\text{diel}})/\partial N$ fulfilling $\int f_{\text{ion+diel}}^{\text{J}}(\mathbf{r}') \text{d}\mathbf{r}' = 1$ as a result of overall charge neutrality. The superscripts ``J'' of the electronic and ionic+dielectric Fukui functions emphasize that they are mutually dependent within JDFT, defined by synchronous variations in both electron and ion numbers.

\begin{figure*}[t]
\includegraphics{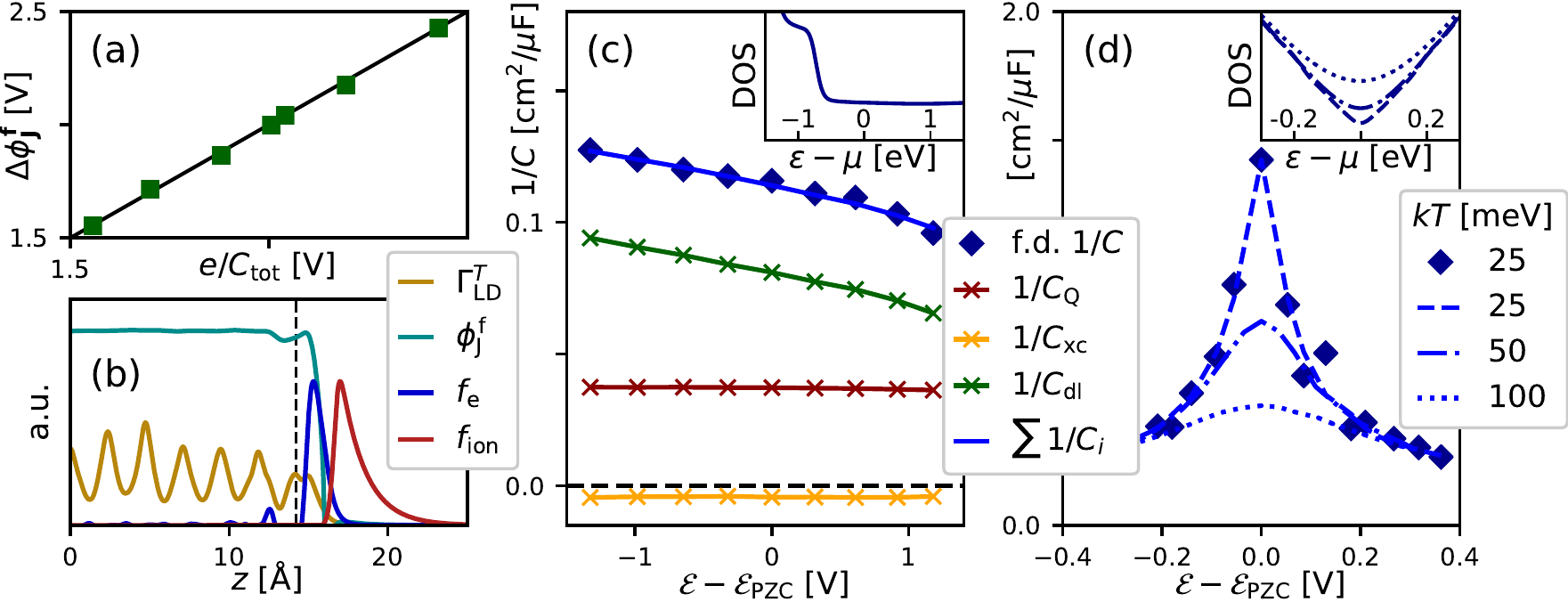}
\caption{JDFT results: (a) and (b) Au (111) slab with 13 layers, $z=0$ at central layer, dashed vertical line through center of surface atoms, $kT=50\,\mathrm{meV}$; the diagonal in (a) corresponds to equality; results in (b) are scaled for improved visibility. (c) Single-layer Au (111) charged from both sides, $kT=50\,\mathrm{meV}$; potential of zero charge $\mathcal{E}_{\text{PZC}}$. (d) Single-layer graphene charged from both sides, $\sum 1/C_i$ at different temperatures. DFT computations with VASP~\cite{1996_CompMaterSci_Kresse}; PAW pseudopotentials~\cite{1999_PhysRevB_Kresse}; PBE XC-functional~\cite{1996_PhysRevLett_PBE}; energy cutoff $450\,\mathrm{eV}$ for Au slabs (a)--(c), and $520\,\mathrm{eV}$ for graphene (d); Fermi-Dirac smearing with $kT$; VASPsol implicit electrolyte model~\cite{2019_JChemPhys_Mathew_Hennig} with $\chi+1=78.4$ and $\lambda_{\text{D}}=1.5\,\mathrm{\AA}$; fixed atomic positions derived from relaxed Au bulk lattice for (a)--(c), and relaxed single-layer graphene in vacuum for (d); $\Gamma$-centered $K$-point meshes: $(27\times 27\times 1)$ for (a)--(b), $(97\times 97\times 1)$ for (c), $(193\times 193\times 1)$ for (d). Filled squares and diamonds in (a) and in (c)--(d), respectively, are computed from finite differences (f.d.), $e^2/C = \Delta\mu/\Delta N$.}
\label{fig_JDFT_results}
\end{figure*}

Eq.~\eqref{eq_hardness_JDFT} is the central result of the present letter, and it represents a rigorous basis for an understanding of the relevant contributions to interface capacitance. Because the shifted electrostatic potential $\phi(\mathbf{r})-\phi_{\text{b}}+\phi_0$ in the KS Hamiltonian is fixed at $\phi_0$ in the bulk electrolyte, the electrode potential is equal to $\mathcal{E}=-\mu/e$ referenced to $\phi_0$ that represents the choice of a certain reference electrode potential. Furthermore, $-e\,\text{d}N=\text{d}Q$ is the change in the charge stored at the electrochemical interface. Therefore, the joint chemical hardness of Eq.~\eqref{eq_hardness_JDFT} is simply the inverse of the total differential interface capacitance per supercell (SC), $\eta_{\text{J}} =\partial\mu/\partial N = e^2\,\partial\mathcal{E}/\partial Q = e^2/C_{\text{tot}}$. The additive nature of Eq.~\eqref{eq_hardness_JDFT} provides a natural interpretation of the total capacitance in terms of a serial circuit of capacitors,
\begin{align}
\frac{1}{C_{\text{tot}}} = \frac{1}{C_{\text{Q}}} + \frac{1}{C_{\text{xc}}} + \frac{1}{C_{\text{dl}}} \ ,
\label{eq_tot_capacitance}
\end{align}
where the quantum capacitance $C_{\text{Q}}=e^2\,g^T_{\text{D}}(\mu)$ is simply the DOS at the Fermi energy, the XC-capacitance $C_{\text{xc}} = e^2/\int \Gamma^T_{\text{LD}}\,\mu^f_{\text{xc}}\text{d}\mathbf{r}$ is the contribution of the Fukui XC-potential, and $C_{\text{dl}}$ is the electrostatic double-layer capacitance given by
\begin{align}
\frac{1}{C_{\text{dl}}} = \frac{1}{e}\left[\phi^f_{\text{J},\text{b}}-\int_{\text{SC}} \phi^f_{\text{J}}(\mathbf{r})\,\Gamma^T_{\text{LD}}(\mathbf{r})\, \text{d}\mathbf{r}\right]\ .
\label{eq_dl_capacitance}
\end{align}
By definition of the joint Fukui potential, $C_{\text{dl}}$ comprises both the electronic excess charge of the electrode, as well as the ionic counter charge layer of the electrolyte. The former can be either a very narrow surface charge layer, as in the case of metal electrodes, or it can be a space-charge layer extending deeply into a semiconductor electrode. 

A partitioning of the total capacitance similar to Eq.~\eqref{eq_tot_capacitance} has been previously used by many authors, e.g.~\cite{2016_JPhysChemLett_Zhan}. To the best of my knowledge, this is the first time that it has been rigorously derived \textit{ab initio}, without including certain contributions \textit{a posteriori}.

For a thick metal electrode, expression~\eqref{eq_tot_capacitance} reduces to a particularly simple form. With a non-zero LDOS at the Fermi energy in the bulk, the total DOS scales with the electrode thickness $t$, so $1/C_{\text{Q}}$ eventually becomes negligible. The integral in Eq.~\eqref{eq_dl_capacitance} corresponds to an average of the Fukui potential weighted by the normalized LDOS. As shown in Fig.~\ref{fig_JDFT_results}b, the latter becomes periodic within the electrode bulk, so the weight of the bulk region turns to one for $t\rightarrow\infty$. Therefore, the integral in Eq.~\eqref{eq_dl_capacitance} converges to the plateau value $\phi^f_{\text{J},(-\text{b})}$ of the Fukui potential inside the electrode bulk (denoted by ``$-\text{b}$''), see Fig.~\ref{fig_JDFT_results}b, and we obtain the simple relation $1/C_{\text{dl}} = \Delta\phi^f_{\text{J}}/e$ with the Galvani step in the joint Fukui potential between electrode and electrolyte. Performing a surface multipole expansion~\cite{2017_PhysRevB_Binninger} of the joint Fukui potential, and noting that the joint Fukui charge is zero, we have $\Delta\phi^f_{\text{J}} = D^{f}_{\text{J}}/\epsilon_0$, where $D^{f}_{\text{J}} = (-e/A_{\text{SC}})\int_{\text{SC}} z\,f_{\text{J}}(\mathbf{r})\,\text{d}\mathbf{r}$ is the average surface dipole of the joint Fukui function, with the surface area $A_{\text{SC}}$ and the surface normal coordinate $z$. With similar arguments, $1/C_{\text{xc}} = \mu^f_{\text{xc},(-\text{b})}/e^2$ is given by the bulk value of the Fukui XC-potential, which we assume to be negligible, because the electronic Fukui function is localized around the metal electrode surface, see Fig.~\ref{fig_JDFT_results}b. Therefore, the total capacitance of a thick metal electrode--electrolyte interface is given by the inverse of the Fukui Galvani potential, or Fukui surface dipole, 
\begin{align}
C_{\text{tot}}\, =\, C_{\text{dl}}\, =\, \frac{e}{\Delta\phi^f_{\text{J}}}\, =\, \frac{\epsilon_0 e}{D^{f}_{\text{J}}}\ , 
\end{align}
as confirmed by the results shown in Fig.~\ref{fig_JDFT_results}a. Note that $D^{f}_{\text{J}} = D^{f}_{\text{e}} + D^{f}_{\text{ion+diel}}$ can be split into an electronic and an electrolyte part, where $D^{f}_{\text{e}} = (-e/A_{\text{SC}})\int_{\text{SC}} z\,f_{\text{e}}^{\text{J}}(\mathbf{r})\,\text{d}\mathbf{r}$ and $D^{f}_{\text{ion+diel}} = (e/A_{\text{SC}})\int_{\text{SC}} z\,f_{\text{ion+diel}}^{\text{J}}(\mathbf{r})\,\text{d}\mathbf{r}$, and accordingly
\begin{align}
\frac{1}{C_{\text{tot}}} = \frac{D^{f}_{\text{e}}}{\epsilon_0 e} + \frac{D^{f}_{\text{ion+diel}}}{\epsilon_0 e} = \frac{1}{C^{f}_{\text{e}}} + \frac{1}{C^{f}_{\text{ion+diel}}} \ .
\label{eq_dl_capacitance_split}
\end{align}
Whereas $D^{f}_{\text{J}}$ is invariant w.r.t. a shift of the $z$-origin, $D^{f}_{\text{e}}$ and $D^{f}_{\text{ion+diel}}$ depend on the choice of the $z=0$ plane, and the splitting~\eqref{eq_dl_capacitance_split} is not unique. It is interesting to note that the effective image plane location $z_{\text{im}}$ of a metal surface is given by the center of mass of the electronic excess charge~\cite{1973_PhysRevB_Lang_Kohn, 1996_ChemRev_Schmickler}, which is essentially equal to the electronic part of the Fukui surface dipole, $z_{\text{im}} = \int_{\text{SC}} z\,f_{\text{e}}^{\text{J}}(\mathbf{r})\,\text{d}\mathbf{r} = (-A_{\text{SC}}/e)D^{f}_{\text{e}}$. Choosing the origin $z_{\text{im}} = 0$ thus corresponds to $D^{f}_{\text{e}}=0$, and the total interface capacitance $C_{\text{tot}} = C^{f}_{\text{ion+diel}}$ is entirely defined by the electrolyte side. The nature of the electrode material still enters via the potential-dependent $z_{\text{im}}$~\cite{2020_JElectroanalChem_Schmickler}. Depending on the details of the electrolyte model, the electrolyte Fukui surface dipole can be further split. If, e.g., there is a Helmholtz layer that is free of ionic charges between the electrode surface at $z_{\text{im}} = 0$ and an outer Helmholtz plane (OHP) at $z_{\text{OHP}}$, then $\int_{z_{\text{OHP}}}^{\infty}\langle f_{\text{ion+diel}}^{\text{J}}\rangle_{A_{\text{SC}}}\text{d}z = 1/A_{\text{SC}}$ because of the overall normalization of the ionic Fukui function ($\langle \cdots\rangle_{A_{\text{SC}}}$ denotes the area average). It follows that $D^{f}_{\text{ion+diel}} = D^{f}_{\text{H}} + D^{f}_{\text{GC}}$, where $D^{f}_{\text{H}} = (e\, z_{\text{OHP}}/A_{\text{SC}}) + \int_{0}^{z_{\text{OHP}}} e\,z\,\langle f_{\text{diel}}^{\text{J}}\rangle_{A_{\text{SC}}}\text{d}z$ and $D^{f}_{\text{GC}} = \int_{z_{\text{OHP}}}^{\infty} e\,(z-z_{\text{OHP}})\,\langle f_{\text{ion+diel}}^{\text{J}}\rangle_{A_{\text{SC}}}\text{d}z$ are the Fukui surface dipoles across the Helmholtz layer and the diffuse Gouy-Chapman layer, respectively. In combination with Eq.~\eqref{eq_dl_capacitance_split} and $D^{f}_{\text{e}}=0$ (due to the choice of origin $z_{\text{im}} = 0$), we obtain the commonly employed spatial partitioning of the electrochemical double-layer capacitance into a Helmholtz capacitance $C_{\text{H}}$ and a Gouy-Chapman capacitance $C_{\text{GC}}$~\cite{1996_ChemRev_Schmickler}, 
\begin{align}
\frac{1}{C_{\text{tot}}} \,=\, \frac{D^{f}_{\text{H}}}{\epsilon_0 e} + \frac{D^{f}_{\text{GC}}}{\epsilon_0 e} \,=\, \frac{1}{C_{\text{H}}} + \frac{1}{C_{\text{GC}}} \ .
\end{align}

The previous considerations for bulk metal electrodes led to a particularly simple form of the total interface capacitance. However, for thin metal electrodes, as well as for materials with space-charge behavior or exotic DOS, the various capacitive contributions are non-trivial and Eqs.~\eqref{eq_hardness_JDFT}--\eqref{eq_dl_capacitance} enable their rigorous quantification. This is exemplified by the computational results presented in Fig.~\ref{fig_JDFT_results} that were obtained using the VASPsol implicit electrolyte model~\cite{2019_JChemPhys_Mathew_Hennig}, where the ionic and dielectric charge densities $\rho_{\text{ion}} = -\epsilon_0 \epsilon_{\text{r}}\kappa^2(\phi-\phi_0)$ and $\rho_{\text{diel}} = \epsilon_0\nabla[\chi\nabla\phi]$ are approximated as linear response to the electrostatic potential and field, respectively, with the electric susceptibility $\chi$ and relative permittivity $\epsilon_{\text{r}} = 1+\chi$ of the electrolyte, and the inverse Debye length $\kappa = 1/\lambda_{\text{D}}$. A single (111) layer of gold has a remarkably flat DOS around the Fermi energy, see inset of Fig.~\ref{fig_JDFT_results}c. The resulting quantum capacitance is almost constant, but it contributes significantly to the total capacitance, see Fig.~\ref{fig_JDFT_results}c. Note that the negative XC-capacitance also comprises the contribution of the electrolyte boundary functional to the local KS-potential in VASPsol~\cite{2019_JChemPhys_Mathew_Hennig}. The slope in $1/C_{\text{dl}}$ results from the electrolyte boundary moving closer to the electrode surface with increasing electrode potential, i.e. decreasing surface electron density. Fig.~\ref{fig_JDFT_results}d shows the inverse total capacitance according to Eq.~\eqref{eq_tot_capacitance} for single-layer graphene at different temperatures. The total capacitance is largely determined by the Dirac cone in the temperature-dependent DOS $g^T_{\text{D}}(\epsilon)$, which becomes softened at increasing temperature, see inset of Fig.~\ref{fig_JDFT_results}d. These results demonstrate how Eqs.~\eqref{eq_hardness_JDFT}--\eqref{eq_tot_capacitance} provide detailed insight into the factors that determine the charging characteristics of complex electrode--electrolyte interfaces, including the influence of the electrode material, thickness, and temperature.

In summary, we resolved the apparent conflict between the piecewise linearity requirement and the concept of capacitance in density functional theory. Within a conceptual JDFT framework, we rigorously derived the commonly employed partitioning of the capacitance of an electrode--electrolyte interface, highlighting the central role of the LDOS and the joint Fukui function, which includes the ionic and dielectric response of the electrolyte. We conclude with an outlook on how these quantities can be calculated. Obviously, the LDOS results directly from a self-consistent JDFT solution of the KS equations~\eqref{eq_KS_electron_density} and~\eqref{eq_KS_Schroedinger} together with, e.g., the linearized Poisson-Boltzmann equation, depending on the electrolyte model employed. The Fukui functions are more complicated to obtain, e.g. from finite differences in $N$. However, since they quantify the linear response of the joint system, they should also be directly computable from a single JDFT solution. In fact, Cohen \textit{et al.}~\cite{1994_JChemPhys_Cohen} showed that the local softness in DFT is given by a linear mapping of the LDOS. Because the DFT Fukui function is equal to the normalized local softness~\cite{1985_PNAS_Yang_Parr}, the former is fully determined by the LDOS, demonstrating the importance of the LDOS in chemical reactivity theory~\cite{1994_PhysRevLett_Cohen}. We foresee a similar relation to hold between the Fukui functions and LDOS in conceptual JDFT. Finally, it should be emphasized that the present analysis is strictly valid for a fixed external potential, i.e. atom core positions. Filhol and Doublet~\cite{2014_JPhysChemC_Filhol} demonstrated how surface structural relaxation contributes to the effective capacitance and can be treated in a conceptual DFT description of surface electrochemistry. Combined with their analysis, the present approach will be generalized to include effects of surface atom displacements and adsorbates.

\begin{acknowledgments}
This work was funded by the SNSF (Swiss National Science Foundation) in the form of a research fellowship grant.
\end{acknowledgments}


%

\end{document}